\UseRawInputEncoding
\documentclass[aps,prl,twocolumn,superscriptaddress]{revtex4-2}

\usepackage{graphicx}  
\usepackage{dcolumn}   
\usepackage{bm}        
\usepackage{amssymb}   

\begin{document}

\newcommand{\sample}{$1T$-$\mathrm{TaS}_2$}

\title{Ultrafast Modulations And Detection of a Ferro-rotational Charge Density Wave Using Time-resolved Electric Quadrupole Second Harmonic Generation}

\author{Xiangpeng Luo}
\affiliation{Department of Physics, University of Michigan, 450 Church St, Ann Arbor, MI 48109, USA}

\author{Dimuthu Obeysekera}
\affiliation{Department of Physics, New Jersey Institute of Technology, 323 Dr Martin Luther King Jr Blvd, Newark, NJ 07102, USA}

\author{Choongjae Won}
\affiliation{Laboratory for Pohang Emergent Materials, Pohang Accelerator Laboratory and Max Plank POSTECH Center for Complex Phase Materials, Pohang University of Science and Technology, Pohang 790-784, Korea}

\author{Suk Hyun Sung}
\author{Noah Schnitzer}
\author{Robert Hovden}
\affiliation{Department of Materials Sciences, University of Michigan, 2300 Hayward Street, Ann Arbor, MI 48109, USA}

\author{Sang-Wook Cheong}
\affiliation{Laboratory for Pohang Emergent Materials, Pohang Accelerator Laboratory and Max Plank POSTECH Center for Complex Phase Materials, Pohang University of Science and Technology, Pohang 790-784, Korea}
\affiliation{Rutgers Center for Emergent Materials and Department of Physics and Astronomy, Rutgers University, Piscataway, NJ 08854, USA}

\author{Junjie Yang}
\affiliation{Department of Physics, New Jersey Institute of Technology, 323 Dr Martin Luther King Jr Blvd, Newark, NJ 07102, USA}

\author{Kai Sun}
\affiliation{Department of Physics, University of Michigan, 450 Church St, Ann Arbor, MI 48109, USA}

\author{Liuyan Zhao}
\email[Corresponding author: ]{lyzhao@umich.edu}
\affiliation{Department of Physics, University of Michigan, 450 Church St, Ann Arbor, MI 48109, USA}

\begin{abstract}
We show the ferro-rotational nature of the commensurate charge density wave (CCDW) in \sample{} and track its dynamic modulations by temperature-dependent and time-resolved electric quadrupole rotation anisotropy-second harmonic generation (EQ RA-SHG), respectively. The ultrafast modulations manifest as the breathing and the rotation of the EQ RA-SHG patterns at three frequencies around the reported single CCDW amplitude mode frequency. A sudden shift of the triplet frequencies and a dramatic increase in the breathing and rotation magnitude further reveal a photo-induced transient CDW phase across a critical pump fluence of $\sim\,0.5\:\mathrm{mJ/cm}^2$.
\end{abstract}

\maketitle

Multipolar orders are key in addressing outstanding questions in a wealth of quantum materials including $f$-electron systems \cite{RN42}, 5$d$ transition metal oxides \cite{RN43}, multiferroics \cite{RN44}, chiral magnets \cite{RN45}, and so on. As the lowest rank multipolar order, the ferro-rotational order, schematically featured as a head-to-tail loop arrangement of electric dipole moments and mathematically described by the antisymmetric components of the electric quadrupole tensor, was theoretically suggested to be widely present \cite{RN37, RN35, RN36}, but has been experimentally detected only very recently with nonlinear optics as linear probes hardly couple with this order \cite{RN25}. While the static physical properties of the ferro-rotational order start to be investigated \cite{RN41, RN25}, its dynamical properties remain unexplored. Electromagnetic (EM) radiation emerges as a promising venue to drive solids into transient states that are dynamical on ultrafast time scales \cite{RN2, RN3, RN4, RN9, RN8, RN10, RN6, RN7, RN5} and/or inaccessible through thermodynamic means \cite{RN15, RN19, RN18, RN20, RN11, RN14, RN17, RN16, RN12, RN50, RN13}, and time-resolved (tr) probes are developed to examine the dynamics of these EM-driven solids. Tr-diffraction-based techniques, e.g., tr-X-ray diffraction, tr-electron diffraction, have been instrumental in capturing the ultrafast evolution of translational symmetries and resolving EM-driven phases and phase transitions, with notable recent examples such as modifying the long-range magnetic orders \cite{RN3, RN9, RN10}, inducing the thermodynamically inaccessible charge orders \cite{RN19, RN50, RN7, RN5}, and driving across the polar phase transitions \cite{RN18, RN17, RN8}. Besides the tremendous success brought by tr-diffraction-based techniques thus far, one may wish to go beyond the translation symmetries particularly sensitive by diffractions and the linear interaction processes dominant in diffraction-based techniques, so as to probe out-of-equilibrium dynamics of unconventional multipolar orders.

Time-resolved ultrafast nonlinear optical spectroscopy complements tr-diffraction-based techniques in detecting subtle point symmetry changes and directly coupling to multipolar order parameters. Second harmonic generation (SHG), the lowest order nonlinear optics, has developed much beyond its conventional practice of detecting non-centrosymmetric crystal structures and managed to reveal structural \cite{RN25}, electronic \cite{RN26}, and magnetic \cite{RN27, RN29, RN28} phase transitions that are characterized by multipolar orders, by performing the high-accuracy rotation anisotropy (RA) measurements \cite{RN30}. So far, the tr-SHG has already shown its power in capturing transient noncentrosymmetric crystal structures by electric dipole SHG \cite{RN31, RN54, RN18, RN17, RN8, RN32}. However, the incorporation of time-resolved capability into the newly developed high-accuracy electric quadrupole (EQ) RA-SHG for detecting multipolar order dynamics awaits to be explored.

\sample{} has a trigonal crystal structure of centrosymmetric point group $D_{3d}$ which consists of one three-fold rotational axis, three diagonal mirrors at every $120^\circ$, and a center of inversion (FIG. 1(a)). Upon cooling, it transits from an incommensurate charge density wave (ICCDW) into a nearly commensurate CDW (NCCDW) phase at $T_\mathrm{NCCDW}=355\:\mathrm{K}$ and then develops into a commensurate CDW (CCDW) phase below $T_\mathrm{CCDW}=183\:\mathrm{K}$ \cite{RN33, RN34}. In addition to the well-known broken translational symmetries for all CDWs, the ICCDW retains all the point symmetries of $D_{3d}$, whereas both the NCCDW and CCDW break the three diagonal mirrors, lowering the symmetry point group from $D_{3d}$ to $S_6$ (FIG. 1(b)). We highlight that, from the perspective of point symmetry, (N)CCDW obeys the ferro-rotational-order-compatible point group $S_6$ and transforms in the same way as the ferro-rotational order, or equivalently, the antisymmetric components of the ferroelectric quadrupolar order \cite{RN37, RN35, RN36}. While the broken translational symmetries of (N)CCDW in \sample{} have been extensive explored by (tr-)diffraction techniques, the mirror-symmetry-broken, spatial-inversion-symmetric ferro-rotational nature of (N)CCDW is a key aspect unfortunately inaccessible previously thus overlooked in literature, and provides an ideal platform to explore the dynamic control and ultrafast detection of this electric quadrupolar order using tr-RA-SHG.

\begin{figure}
\includegraphics[width=\columnwidth]{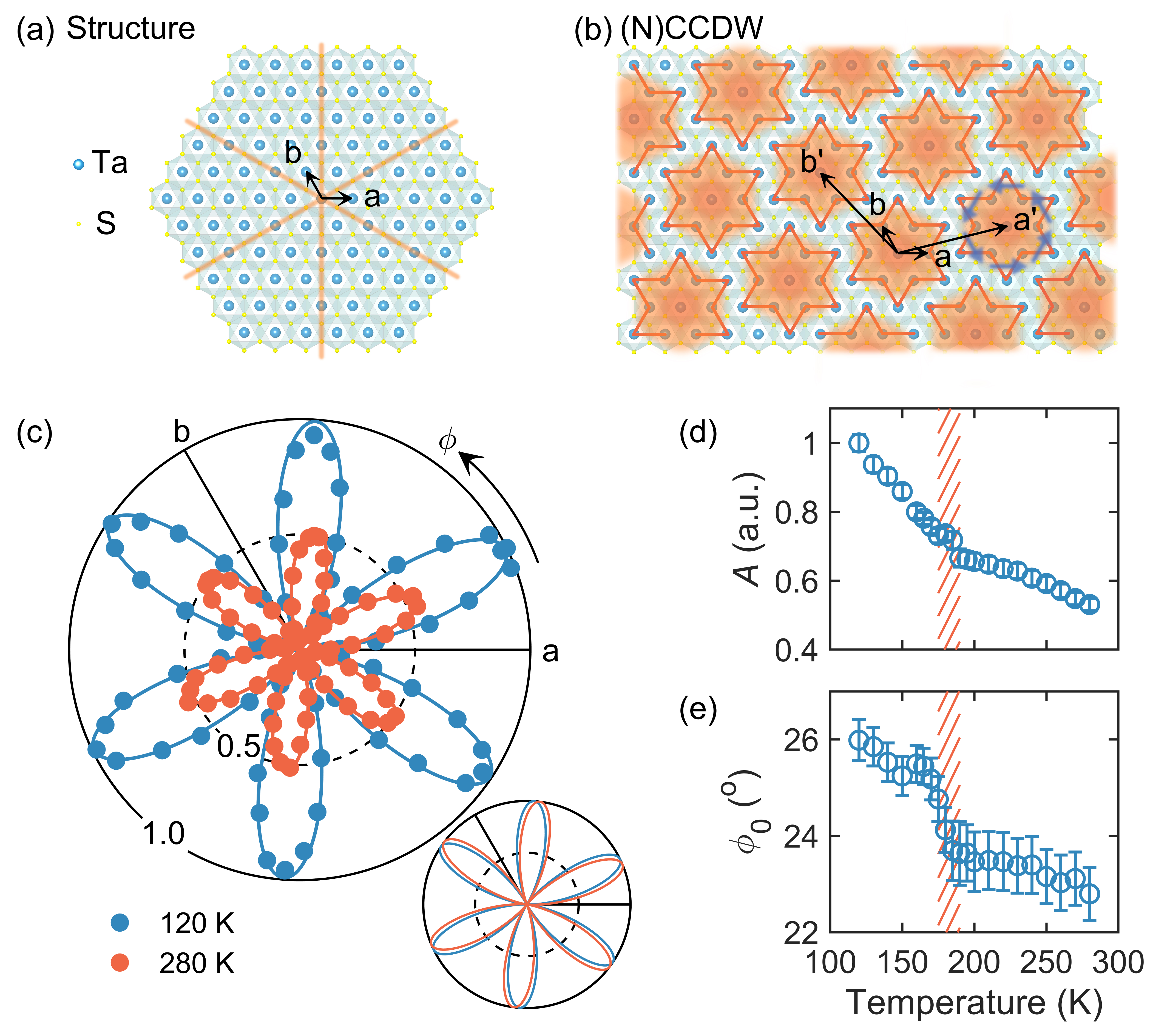}%
\caption{\label{fig1}(a) The top view of \sample{} lattice structure in the a-b plane, with crystal axes $\bm a$ and $\bm b$ marked with black arrows. Three orange lines represent the three diagonal mirrors in the $D_{3d}$ point group. (b) The sketch of the ordering of star-of-David clusters in the (N)CCDW phase. The $\sqrt{13}\times\sqrt{13}$ superlattice axes $\bm a'$ and $\bm b'$ rotate away from the crystal axes $\bm a$ and $\bm b$, and therefore the (N)CCDW breaks the three diagonal mirrors in (a), lowering the symmetry point group into the ferro-rotational $S_6$. The head-to-tail arrangement of blue arrows sketches the ferro-rotational order. (c) Polar plots of RA-SHG data taken at 120 K in the CCDW phase (blue) and at 280 K in the NCCDW phase (orange). Solid lines are fits to the calculated EQ RA-SHG functional form $I^{2\omega}_\perp = A\cos^2 3(\varphi-\varphi_0)$. Inset shows the 120 K and 280 K fits normalized to their own maxima values to illustrate the increased rotation at the lower temperature. (d-e) Temperature dependence of RA-SHG pattern amplitude $A$ (d) and angle of rotation from $\bm a$ axis $\varphi_0$ (e) in a cooling cycle. The transition from the NCCDW to CCDW phase is captured at around 185 K (marked as orange vertical strips). Error bars stand for one standard error of the fits to extract $A$ and $\varphi_0$.}
\end{figure}

We start with establishing the ferro-rotational nature of (N)CCDW in \sample{} by performing temperature dependent static RA-SHG measurements across $T_\mathrm{CCDW}$ \cite{RN46}\nocite{RN53, RN47, RN48}. Figure 1(c) main panel shows the polar plots of RA-SHG data taken at 280 K and 120 K, above and below $T_\mathrm{CCDW}$ respectively, in the crossed polarization channel at the normal incidence on the \sample{} layers, noted as $I^{2\omega}_{\perp}(\varphi)$, where $\varphi$ is the angle between the crystal axis $\bm a$ and the incident fundamental polarization while the reflected SHG is selected by an analyzer always perpendicular to the incident polarization. It is apparent that both patterns rotate away from the crystal axis $\bm a$, evidencing broken diagonal mirrors that are prescribed by the ferro-rotational point group $S_6$ \cite{RN46}. The symmetry selection dictates that the RA-SHG takes the functional form $I^{2\omega}_\perp = A\cos^2 3(\varphi-\varphi_0)$, where $A$ and $\varphi_0$ stand for the amplitude and orientation of the RA-SHG pattern, respectively and relate to the bulk EQ susceptibility tensor $\chi_{ijkl}^\mathrm{EQ}$ via $A=\sqrt{\chi_{xxzx}^2+\chi_{yyzy}^2}$ and $\textstyle\varphi_0=\frac{1}{3}\mathrm{atan}\frac{\chi_{xxzx}}{\chi_{yyzy}}$ under the centrosymmetric point group $S_6$, with the surface effect convincingly ruled out by thickness-dependent RA-SHG measurements \cite{RN46}. The experimental RA-SHG data at both temperatures fit well to the simulated functions (solid lines in FIG. 1(c) main panel), and the fitted results show a clear enhancement in the amplitude A and a resolvable increase in the orientation $\varphi_0$ at the lower temperature that are visible in the main panel and inset of FIG. 1(c) respectively. Furthermore, the thorough temperature dependence of EQ RA-SHG retains the three-fold rotational symmetry between 280 and 120 K, showing no signature of triclinic CDW observed in the early study {\cite{RN55}}, whereas that of $A$ shows a kink of changing slopes across $T_\mathrm{CCDW}$ and that of $\varphi_0$ exhibits a jump at $T_\mathrm{CCDW}$, as shown in FIG. 1(d) and 1(e), respectively. Both behaviors are consistent with the enhancement of the ferro-rotational order parameter across the transition from NCCDW with short-range ordered star-of-David patches to CCDW with long-range ordered uniform domains \cite{RN38}, where a rotation of superlattice wavevectors by $\sim2^\circ$ was observed by diffraction techniques {\cite{RN33}}. 

\begin{figure}
\includegraphics[width=\columnwidth]{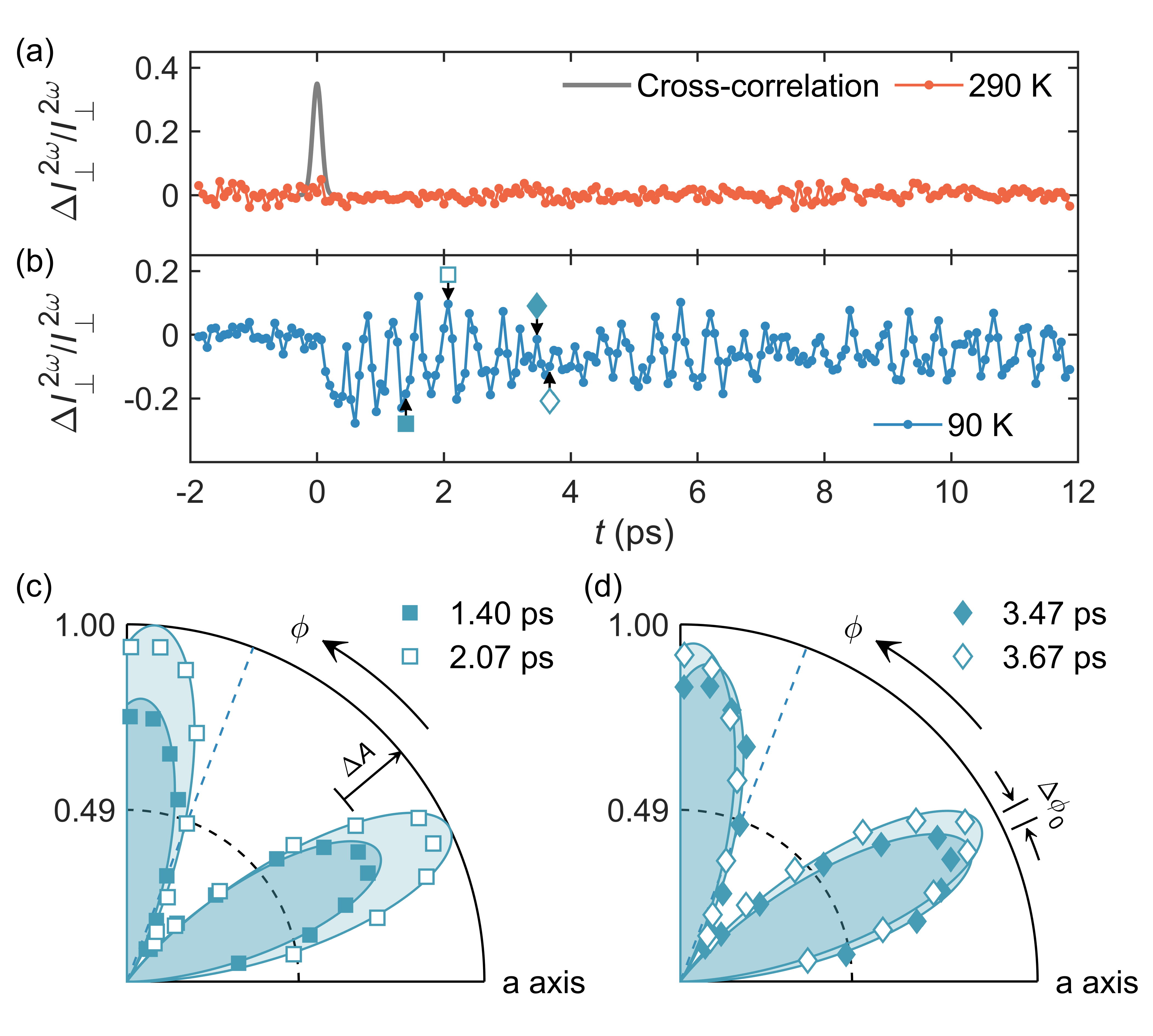}%
\caption{\label{fig2}(a) Normalized tr-SHG intensity (orange) at $\varphi=69.3^\circ$ taken at 290 K in the NCCDW phase. the cross-correlation function (gray) between pump and probe pulses is shown to mark time-zero and time resolution. (b) Normalized tr-SHG trace (blue) at $\varphi=69.3^\circ$ taken at 90 K in the CCDW phase, showing both the slow recovery process after the sudden suppression at time-zero and the fast coherent oscillations with a beating profile. (c-d) Comparisons between transient RA-SHG patterns at two pairs of delay time marked by black arrows in (b). The dashed blue radial line represents the polarization angle $\varphi=69.3^\circ$ at which the tr-SHG trace in (b) was measured. The dashed black arc at 0.49 represents the pre-time zero SHG intensity level in (b). The comparisons in (c) and (d) demonstrate the change in SHG present in both the RA-SHG amplitude and orientation channels.}
\end{figure}

We are now ready to explore the dynamics of the ferro-rotational order by pumping the NCCDW and CCDW phases with an optical pulse at 720 nm that creates a transient imbalance of photo-induced electrons and holes \mbox{\cite{RN12, RN50}} and probing the ultrafast evolution of both phases with the SHG intensity in the crossed channel at $\varphi=69.3^\circ$. Figures 2(a) and 2(b) show the plots of relative change in the SHG intensity, $\frac{\Delta I^{2\omega}_\perp(t)}{I^{2\omega}_\perp(t<0)}$ with $\Delta I^{2\omega}_\perp(t)=I^{2\omega}_\perp(t)-I^{2\omega}_\perp(t<0)$, taken at 290 K for NCCDW and 90 K for CCDW. The time-zero $t=0\:\mathrm{ps}$ and the time resolution of $t_\mathrm{res}=0.09\:\mathrm{ps}$ are determined by the peak and the full-width-of-half-maximum of the cross-correlation function between the pump and probe pulses (gray solid line in FIG. 2(a)). In contrast to the absence of any time-dependent change for NCCDW (FIG. 2(a)), which is expected due to strong suppression and damping of excitations in short-range orders \mbox{\cite{RN56, RN38}}, the tr-SHG trace for CCDW shows two prominent features: the slow recovery after the sudden suppression at time-zero and the fast coherent oscillations with a beating profile (FIG. 2(b)). To identify the sources of both time-dependent features, we select two pairs of delay time (marked by black arrows in FIG. 2(b)) and compare the RA-SHG polar plots within each pair. The two RA-SHG patterns at $t=1.40\:\mathrm{ps}$ and $2.07\:\mathrm{ps}$ show a clear change in the amplitude ($\Delta A$ as marked in FIG. 2(c)), whereas those at $t=3.47\:\mathrm{ps}$ and $3.67\:\mathrm{ps}$ display a notable change in the orientation ($\Delta\varphi_0$ as marked in FIG. 2(d)). This observation makes sense as the static ferro-rotational order parameter of CCDW is indeed encoded in both the amplitude and orientation of RA-SHG patterns (FIG. 1(c-e)). Therefore, we need to track the time-dependence of both the amplitude and the orientation channels, $\Delta A(t)$ and $\Delta\varphi_0(t)$, in order to get a comprehensive understanding on the dynamic modulations of this ferro-rotational CCDW.

We hence proceed to conduct the angle $\varphi$ dependent RA-SHG measurements at every single delay time t and construct the map of tr-RA-SHG---the SHG intensity $I^{2\omega}_\perp$ as functions of $t$ and $\varphi$---taken in the crossed polarization channel at 90 K (FIG. 3(a)) \cite{RN46}. In this map, a horizontal slice is a tr-SHG trace akin to FIG. 2(b), and a vertical cut is a transient RA-SHG pattern in analogy to FIG. 1(c). We fit the individual RA-SHG pattern at every delay time $t$ with $I^{2\omega}_\perp(\varphi,t)=A(t)\cos^2 3(\varphi-\varphi_0(t))$ and obtain the time dependence of both amplitude and orientation, $A(t)$ and $\varphi_0(t)$. Their changes with respect to the pre-time-zero $(t<0)$ values are plotted in FIG. 3(b) after a normalization, $\frac{\Delta A(t)}{A(t<0)}$, and in FIG. 3(c) in absolute size, $\Delta\varphi_0(t)$, respectively. Comparing the two traces, we notice two notable differences. First, it is evidently present in the amplitude but absent in the orientation channel that a sudden suppression of signal happens right upon the pump excitation (i.e., $t=0$) and recovers slowly over a couple of picoseconds. Second, the beating profiles of the fast coherent oscillations apparently show distinct phases and frequencies between the two traces, which corroborates with the difference in their fast Fourier transformation (FFT) spectra in FIG. 3(d). 

\begin{figure*}
\includegraphics[width=\textwidth]{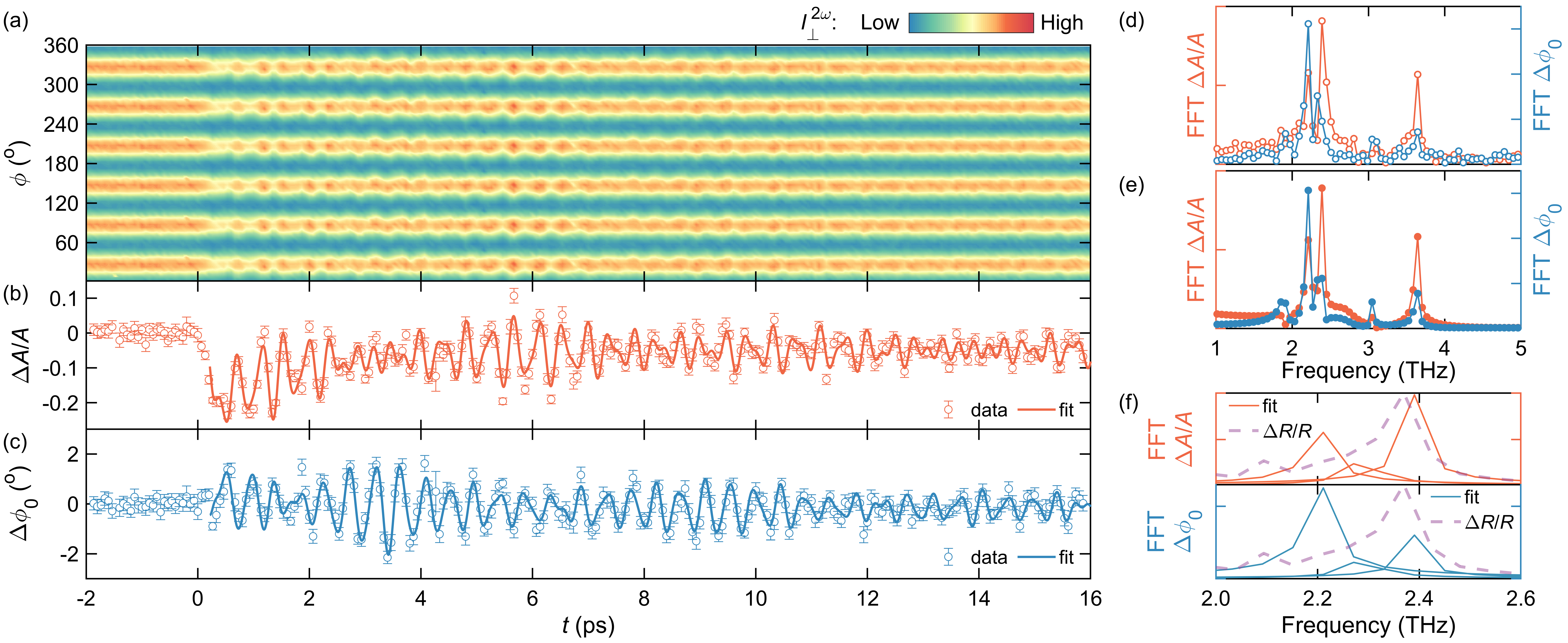}%
\caption{\label{fig3}(a) Map of tr-RA-SHG intensity as functions of the polarization $\varphi$ and the delay time $t$ taken at 90 K. (b-c) Time dependent changes of amplitude normalized to the pre-time-zero value (b) and orientation in absolute size (c), after fitting RA-SHG at individual delay time. Solid lines show the fits of both traces to a functional form consisting of one exponential decay background $M_\mathrm{B}e^{-t⁄\tau_\mathrm{B}}$ and six under-damped oscillations $\sum_{i=1}^6 M_i e^{-t/\tau_i}\cos(\omega_i t+\delta_i)$. Error bars stand for one standard error in fitting RA-SHG data. (d-e) FFTs of the raw traces for time dependent changes in the amplitude and orientation channels (d) and of the fits for them (e). (f) Zoom-in plots of FFTs of the fitted traces near the CCDW amplitude mode frequency for the amplitude (orange) and orientation (blue) channels, detailing the triplet fine structure. The FFTs of tr-fundamental reflectivity ($\Delta R/R$) (purple) is plotted as a comparison.}
\end{figure*}

To better quantify the differences between the dynamics in the amplitude and the orientation channels, we fit the $\frac{\Delta A(t)}{A(t<0)}$ and $\Delta\varphi_0(t)$ traces simultaneously with one exponential decay background $M_\mathrm{B}e^{-t⁄\tau_\mathrm{B}}$ and at least six under-damped oscillations $\sum_{i=1}^6 M_i e^{-t/\tau_i}\cos(\omega_i t+\delta_i)$, decided by the collection of tr-RA-SHG at all pump fluences (see later). Here, the decay time constants, $\tau_\mathrm{B}$ and $\tau_i$, and the oscillation frequencies $\omega_i$ are kept the same for both traces, whereas the magnitudes, $M_\mathrm{B}$ and $M_i$, and the oscillation phases $\delta_i$ are allowed to vary between the two traces. The fitted lines (solid lines in Figures 3(b) and 3(c)) well capture all key features of $\frac{\Delta A(t)}{A(t<0)}$ and $\Delta\varphi_0(t)$ traces, and the FFTs of fitted traces in FIG. 3(e) nicely reproduce those of the raw spectra in FIG. 3(d). For the slow incoherent recovery process, indeed it only has a finite magnitude in the amplitude channel with a decay constant of $\tau_\mathrm{B}=1.6\:\mathrm{ps}$ but little magnitude in the orientation channel. For the fast coherent oscillations, six distinguishable frequencies are identified in both channels, three of which at 1.89$\pm$0.02, 3.06$\pm$0.01 and 3.628$\pm$0.006 THz are phonon modes observed in Raman spectra \cite{RN39} and the other three at 2.201$\pm$0.004, 2.29$\pm$0.01, and 2.387$\pm$0.005 THz are around the CCDW amplitude mode reported by both Raman \cite{RN39} and time-resolved reflectivity \cite{RN40} measurements. Of particular interest is the latter triplet, because of its close tie to the ferro-rotational CCDW and its higher spectral weight than the rest. The individual modes of this triplet in the amplitude and the orientation channels are shown in FIG. 3(f). It is worth noting that our tr-RA-SHG data reveals that the single amplitude mode seen in tr-fundamental reflectivity spectra \cite{RN46} in fact contains a nontrivial triplet structure with distinct distributions in the amplitude and orientation channels. This triplet structure is likely to result from the multiple collective excitations of the ferro-rotational CCDW, which in principle may include the amplitude mode similar to that of conventional CDWs, the rotational and the breathing modes related to the ferro-rotational aspect of this CCDW,  although future studies are needed to assign individual modes in the triplet to specific excitations.

Having established tr-RA-SHG probing the complex dynamics of the ferro-rotational CCDW in \sample{}, we finally proceed to explore its pump fluence dependence. We carried out the maps of tr-RA-SHG and performed the same analysis procedure as in Fig. 3 at five different pump fluences, 0.36, 0.46, 0.58, 0.66 and 0.92 $\mathrm{mJ/cm}^2$. The extracted $\frac{\Delta A(t)}{A(t<0)}$ and $\Delta\varphi_0(t)$ traces and their fits are shown in FIG. 4(a) and 4(b), respectively. As the fast coherent oscillations dominate the slow incoherent recovery process in the amplitude channel at high fluences (FIG. 4(a)), as well as that the incoherent recovery time constant shows little fluence dependence, we focus our discussion on the coherent process. We show the fitted triplet frequencies near the CCDW amplitude mode in FIG. 4(c) and the strongest magnitude in the orientation channel in FIG. 4(d). First, the frequencies of the triplet fine structure show a sudden shift at a critical fluence of $F_C\sim0.5\:\mathrm{mJ/cm}^2$, which contributes to the beating profile evolution upon increasing the pump fluence in FIG. 4(a) and 4(b). In contrast, such a frequency shift is not detectable in the tr-fundamental reflectivity (FIG. 4(c), purple diamonds). Second, the magnitude of the oscillations experiences a dramatic increase across $F_C$ that is clearly visible in traces in FIG. 4(b). Both anomalous behaviors of frequency shift and magnitude enhancement across $F_C$ are indicative of a potential EM radiation-induced phase transition. Considering the fact that there is no tr-SHG signal observed in the NCCDW phase at $T>T_\mathrm{CCDW}$ whereas clear dynamics in tr-SHG is present for all fluences investigated, we can confidently rule out the possibility of this observed photo-induced phase transition resulting from a photo-heating-induced transition from CCDW into NCCDW. In fact, the photo-heating effect from the pump in this study is minimal as there is no pre-time-zero changes in tr-reflectivity and tr-RA-SHG observed even at our highest fluence {\cite{RN46}}. We thus attribute $F_C$ as the critical point where the pump-induced electron-hole imbalance is big enough to destroy the star-of-David clusters and lead to a transient new CDW phase.

\begin{figure}
\includegraphics[width=\columnwidth]{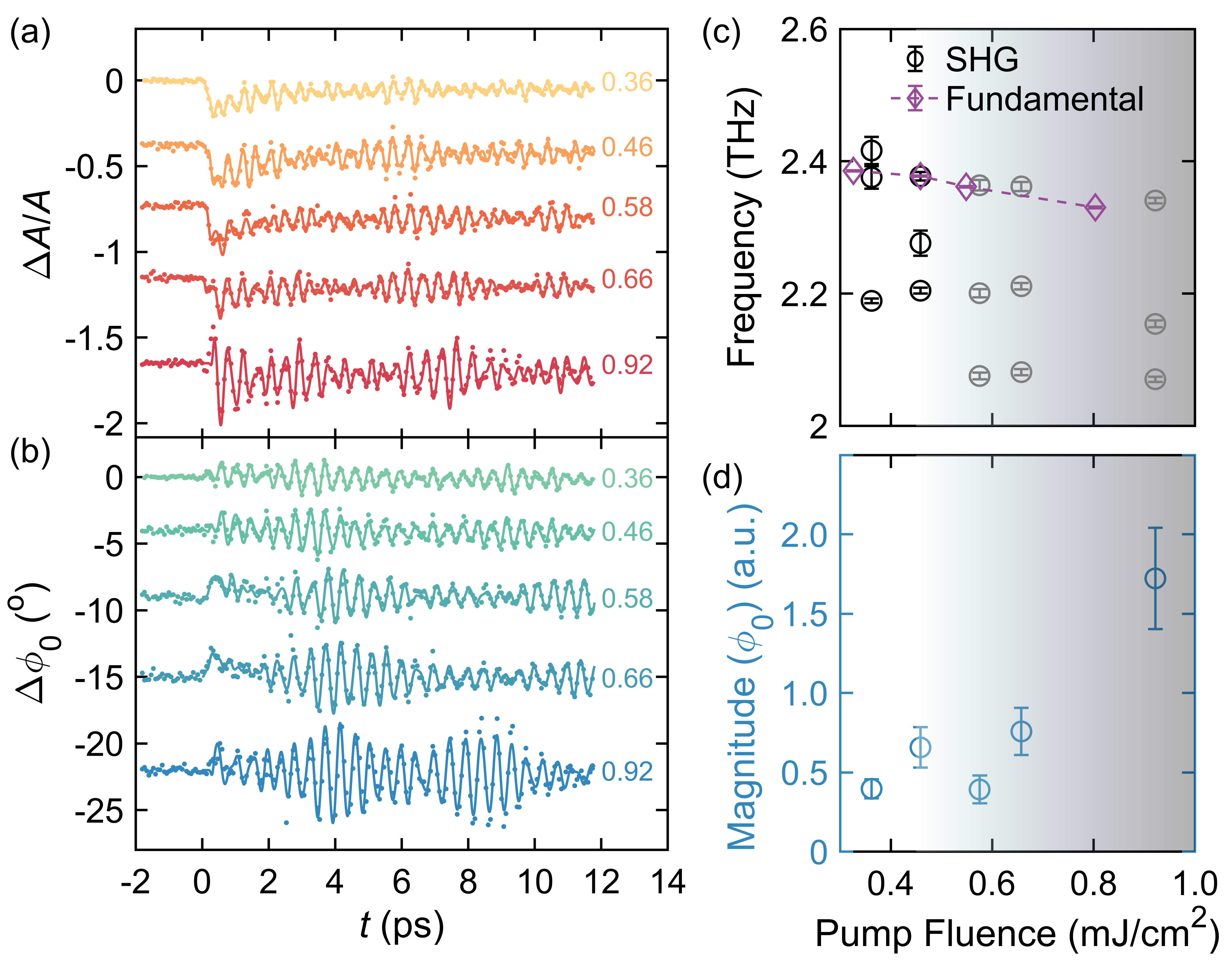}%
\caption{\label{fig4}(a-b) Fluence dependence of the time dependent amplitude (a) and orientation (b) changes fitted from tr-RA-SHG maps taken at pump fluences of 0.36, 0.46, 0.58, 0.66 and 0.92 mJ/cm$^2$. Solid lines are fits to the same functional form in Figure 3. (c) The fluence dependence of the fitted frequencies for the triplet structure around the CCDW amplitude mode (black and gray circles) in tr-RA-SHG and that of the CCDW amplitude mode (purple diamonds) in tr-fundamental reflectivity.  (d) The fluence dependence of the strongest magnitude in the triplet in the orientation channel. The gradient background marks the frequency shift in (c) and the magnitude enhancement in (d) across a critical pump fluence of 0.5 mJ/cm$^2$. Error bars represent one standard error in the fits.}
\end{figure}

Finally, we would like to discuss the comparison between this photo-induced CDW phase that is transient, short-lived and the previously reported optically manipulated CDWs in \sample{}. First, a single femtosecond light pulse with an incident fluence of $\geqslant5\:\mathrm{mJ/cm}^2$ at room temperature was reported to create or destroy metastable mirror-related domains in the NCCDW state \cite{RN5}. In addition to our much lower critical fluence of 0.5 $\mathrm{mJ/cm}^2$ and much shorter lifetime of a few picoseconds, the observed collective mode frequency shift above $F_C$ is not compatible with creation or annihilation of energetically degenerated, mirror-symmetry-related CCDW domains because they should host the very same collective excitations as the CCDW phase below $F_C$. Second, a single 35 fs, $\geqslant 1\:\mathrm{mJ/cm}^2$ light pulse below 70 K was shown to induce a hidden metastable, metallic CDW phase whose amplitude mode {\cite{RN12}} and wavevector {\cite{RN50}} are slightly shifted from those of CCDW. Further studies on this photo-induced hidden metastable CDW phase revealed that its lifetime significantly decreases at higher temperatures \mbox{\cite{RN52, RN51}}. Considering the similar frequency shift but higher temperature and lower critical fluence in our study compared to those in literature, it is likely that our photo-induced transient CDW here is a short-lived version of the hidden metastable CDW phase, which is observable at the ultrafast timescale thanks to the direct coupling between tr-EQ-RA-SHG and the ferro-rotational nature of this CDW.

\begin{acknowledgments}
L. Zhao acknowledges support by NSF CAREER Grant No. DMR-174774 and AFOSR Grant No. FA9550-21-1-0065. K. Sun acknowledges support by NSF Grant No. NSF-EFMA-1741618. S.-W. Cheong acknowledges that the work at Rutgers University was supported by the NSF under Grant No. DMR-1629059 and the work at Postech was supported by the National Research Foundation of Korea (NRF) funded by the Ministry of Science and ICT (No. 2016K1A4A4A01922028). R. Hovden acknowledges support from W. M. Keck Foundation.
\end{acknowledgments}

\end{document}


\newcommand{\sample}{$1T$-$\mathrm{TaS}_2$}
\renewcommand{\thefigure}{S\arabic{figure}}
\renewcommand{\theequation}{S\arabic{equation}}


\title{Supplemental Material for ``Ultrafast Modulations And Detection of a Ferro-rotational Charge Density Wave Using Time-resolved Electric Quadrupole Second Harmonic Generation''}


\author{Xiangpeng Luo}
\affiliation{Department of Physics, University of Michigan, 450 Church St, Ann Arbor, MI 48109, USA}

\author{Dimuthu Obeysekera}
\affiliation{Department of Physics, New Jersey Institute of Technology, 323 Dr Martin Luther King Jr Blvd, Newark, NJ 07102, USA}

\author{Choongjae Won}
\affiliation{Laboratory for Pohang Emergent Materials, Pohang Accelerator Laboratory and Max Plank POSTECH Center for Complex Phase Materials, Pohang University of Science and Technology, Pohang 790-784, Korea}

\author{Suk Hyun Sung}
\author{Noah Schnitzer}
\author{Robert Hovden}
\affiliation{Department of Materials Sciences, University of Michigan, 2300 Hayward Street, Ann Arbor, MI 48109, USA}

\author{Sang-Wook Cheong}
\affiliation{Laboratory for Pohang Emergent Materials, Pohang Accelerator Laboratory and Max Plank POSTECH Center for Complex Phase Materials, Pohang University of Science and Technology, Pohang 790-784, Korea}
\affiliation{Rutgers Center for Emergent Materials and Department of Physics and Astronomy, Rutgers University, Piscataway, NJ 08854, USA}

\author{Junjie Yang}
\affiliation{Department of Physics, New Jersey Institute of Technology, 323 Dr Martin Luther King Jr Blvd, Newark, NJ 07102, USA}

\author{Kai Sun}
\affiliation{Department of Physics, University of Michigan, 450 Church St, Ann Arbor, MI 48109, USA}

\author{Liuyan Zhao}
\email[Corresponding author: ]{lyzhao@umich.edu}
\affiliation{Department of Physics, University of Michigan, 450 Church St, Ann Arbor, MI 48109, USA}



\maketitle

\subsection{Growth of \bm{$1T$}\text{-}\bm{$\mathrm{TaS}_2$} single crystals}

Single crystals of \sample{} were grown via chemical vapor transport (CVT) in Iodine (I$_2$). High purity powders of Ta and S were first mixed in a 1:2 stochiometric ratio and then were finely grounded and pelletized. The pellets were sealed in quartz tubes and sintered at 750 $^\circ\mathrm{C}$ for 24 hours. Afterwards, the pellets were grounded and pelletized again, and the same sintering process was repeated to obtain the polycrystalline power samples. The polycrystalline powder was then sealed in a 200 mm long quartz tube with I$_2$ as the transport agent. The tube was placed in a two-zone tube furnace with the hot end at 1000 $^\circ\mathrm{C}$ and the cool end at 900 $^\circ\mathrm{C}$ and kept at this condition for one week before quenched in water. Large thin hexagonal crystals were finally obtained at the cool end of the tube.

\subsection{Electron microscopy}

Crystallographic orientations were identified by selected area electron diffraction (SAED) on JEOL 2010F and convergent beam electron diffraction (CBED) on JEOL 3100R05 transmission electron microscope, operated at 200 keV and 300 keV, respectively. Rotational calibration between real space image and diffraction pattern were performed by aligning shadow-grams---formed by diverging Bragg spots on SAED and defocusing CBED patterns---to the real space image.

\subsection{RA-SHG, tr-SHG, and tr-RA-SHG measurements}

In the RA-SHG measurements with the normal incidence geometry, the reflected SHG intensity is recorded as a function of the azimuthal angle ϕ between the incident electric polarization and the in-plane crystalline axis $\bm a$, with the reflected electric polarization 0$^\circ$ (90$^\circ$) from the incident one for the parallel (crossed) polarization channel. In this experiment, the incident ultrafast light source was of 800 nm wavelength, 50 fs pulse duration and 200 kHz repetition rate, and was focused on to a 30 $\mu$m diameter spot on the sample with a fluence of $\sim$ 0.5 mJ/cm$^2$. The intensity of the reflected SHG was measured with a single photon counting detector. In the tr-SHG measurements, a pump beam is added in the collinear geometry with the probe beam. The ultrafast light source was of 720 nm wavelength, 75 fs pulse duration and 200 kHz repetition rate, and was focused on to a 65 $\mu$m diameter spot on the sample with a tunable fluence from 0.3 to 1.0 mJ/cm$^2$. The probe beam performs the SHG measurement at a selected polarization angle $\varphi=\varphi_S$ as a function of the time delay between the pump and probe pulses. In the tr-RA-SHG measurements, the pump beam remains the same as that in tr-SHG whereas the probe beam performs RA-SHG measurements at every single time delay.

\subsection{Determination of crystallographic axes and broken mirror symmetries in the NCCDW}

The RA-SHG pattern breaking mirror symmetry under the ferro-rotational point group $S_6$ of the NCCDW phase is confirmed by comparing the orientation of RA-SHG pattern with respect to the crystal axes that were determined by SAED and CBED. As shown in FIG. S1, SAED \cite{RN1} together with RA-SHG was taken on a same thin flake of \sample{}. The RA-SHG pattern in FIG. S1(c), taken in the crossed polarization channel, is plotted with respect to the crystal axes $\bm a$ and $\bm a$ that are calibrated through the CBED measurement (FIG. S1(b) right panel). The observation of the RA-SHG pattern rotating away from the crystal axis $\bm a$ by 22.8$^\circ$ (shaded in blue) confirms that the NCCDW phase breaks the three diagonal mirrors at 30$^\circ$o, 90$^\circ$ and 150$^\circ$ from the axis $\bm a$.

\begin{figure}
\includegraphics[width=\columnwidth]{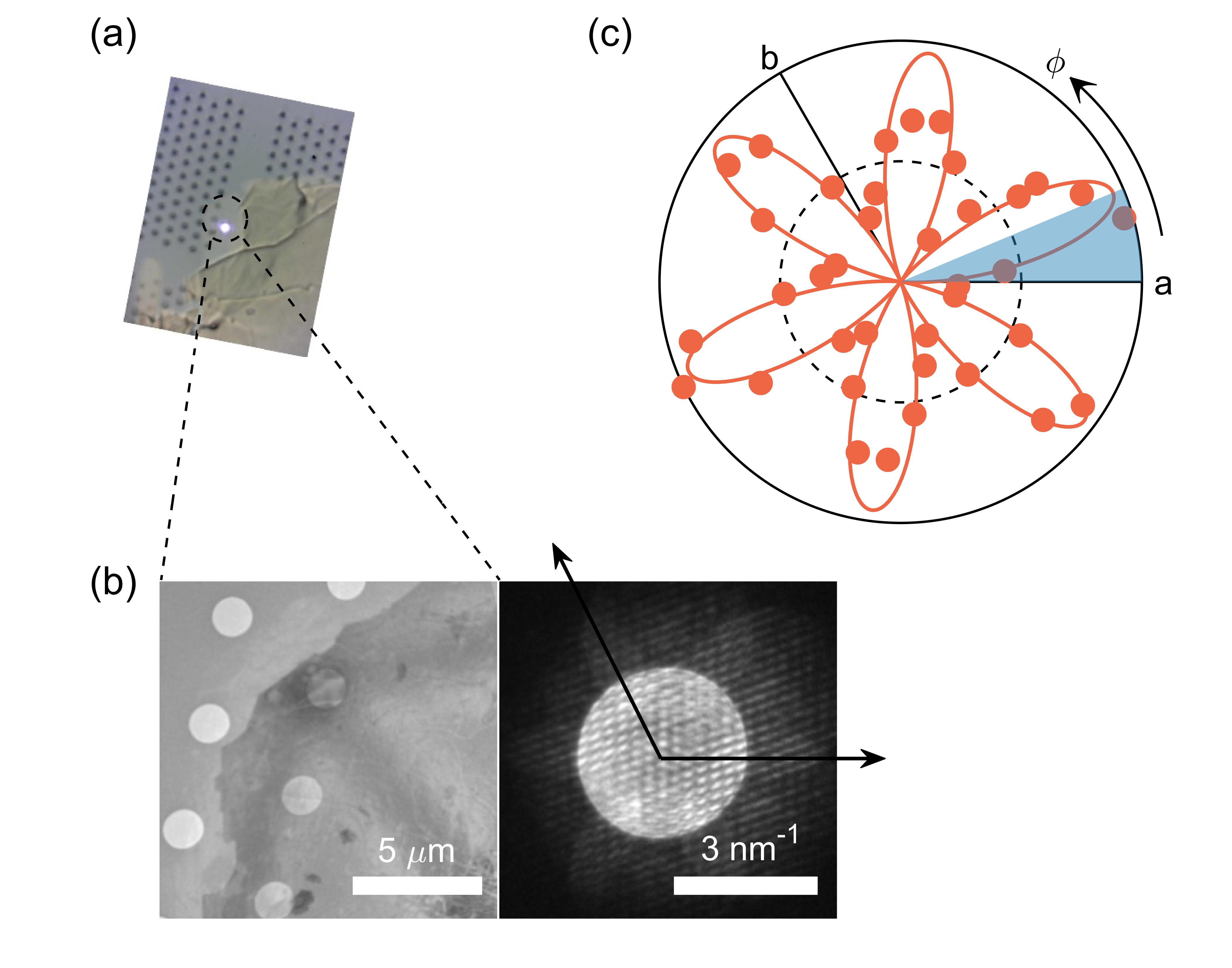}%
\caption{(a) A thin flake of \sample{} placed on a transmission electron microscopy grid. The bright spot is a focused laser beam for performing the RA-SHG measurements in (c). (b) Left: the transmission electron microscopy image of the flake in part. Right: the CBED pattern showing the Bragg disks from the diffracted diverging electrons when the sample is illuminated by a convergent electron beam. Arrows denote the crystal axes. (c) Polar plot of RA-SHG data taken at 290 K with the normal incidence geometry on the (001) plane and in the crossed polarization channel. The solid line is the fit to the simulated RA-SHG functional form $I^{2\omega}_\perp=A\cos^2 3(\varphi-\varphi_0)$. The blue shade represents the rotation of the RA-SHG from the crystal axis $\bm a$.}
\end{figure}

\subsection{RA-SHG measurements on bulk and film \sample{} samples}

RA-SHG measurements are conducted on both bulk and film (less than 100 nm thick) \sample{} samples at room temperature, with fluences of $\sim$ 0.5 and 0.4 mJ/cm$^2$, respectively. Figure S2 shows the RA-SHG patterns taken at the crossed polarization channel, both normalized with respect to incident fluence and integration time. Noting the penetration depth in \sample{} at 800 nm is $\sim$ 28 nm \cite{RN4}, the significant decrease of SHG intensity from bulk to film samples strongly suggests the bulk origin of the SHG signal, and therefore the assigned $S_6$ point group as well as the EQ process generating the SHG.

\begin{figure}
\includegraphics[width=0.8\columnwidth]{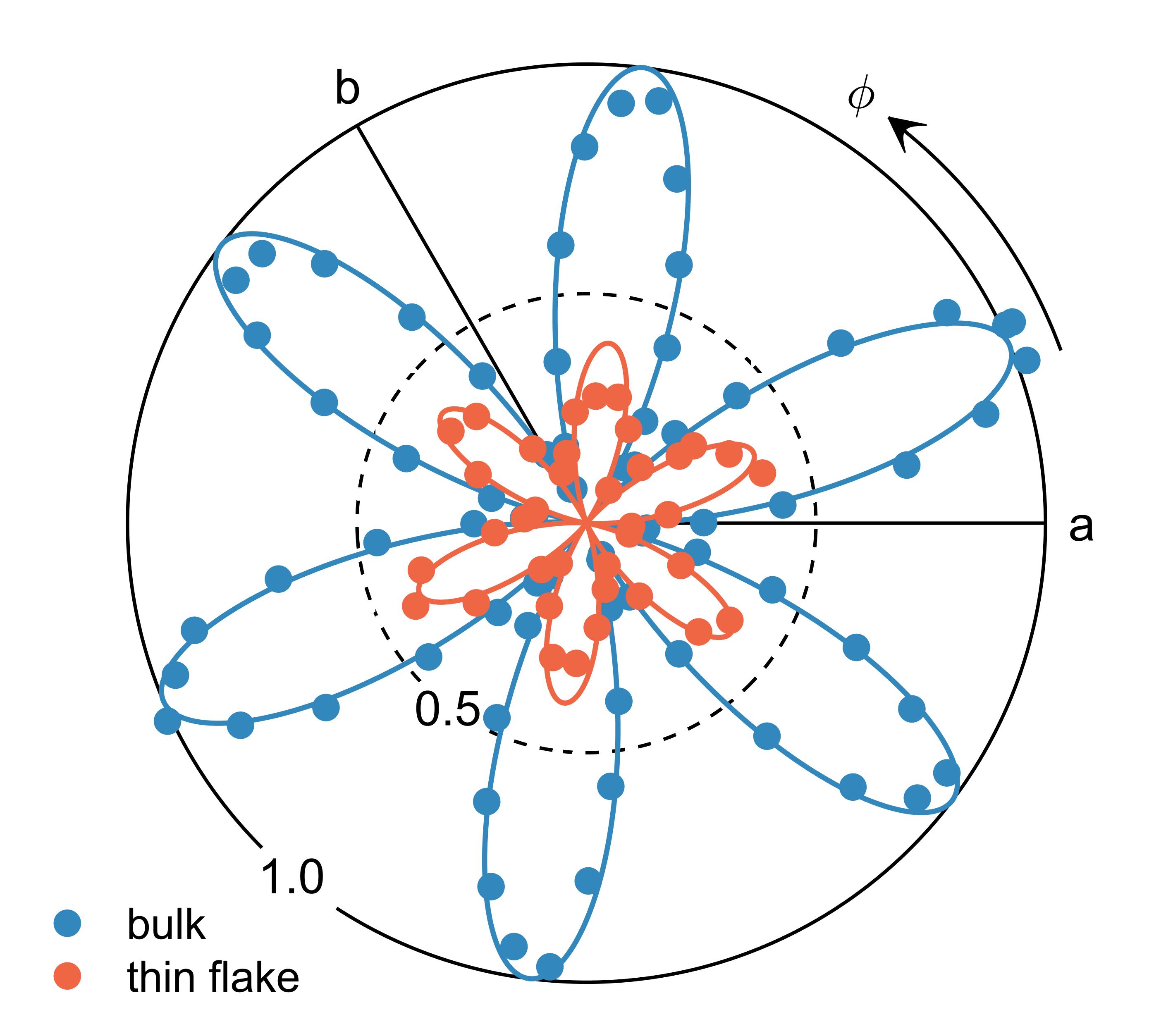}%
\caption{Polar plots of RA-SHG data taken on bulk (blue) and thin flake (orange) \sample{} samples at room temperature.}
\end{figure}

\subsection{Simulations of RA-SHG fitting functions}

In the (N)CCDW phase of \sample{}, the system possesses the point group S6 where spatial inversion is preserved. The observed SHG is dominated by the bulk electric quadrupole (EQ) term \cite{RN2}
\begin{equation}\label{EQRadiation}
    P^\mathrm{eff}_i (2\omega) = \chi^\mathrm{EQ}_{ijkl}E_j(\omega)\partial_k E_l(\omega)
\end{equation}
where $P^\mathrm{eff}_i (2\omega)$ is the effective radiation source for the SHG, $\chi^\mathrm{EQ}_{ijkl}$ is the second-order EQ optical susceptibility tensor under the $S_6$ point group, $E_{j,l}(\omega)$ is the incident fundamental electric field, and $\partial_k$ turns into the wavevector $q_k$ of the incident beam. The RA-SHG patterns can be fitted to the expression
\begin{equation}\label{SHGInt}
    I^{2\omega}\propto |\bm{e}(2\omega)\cdot\bm{P}^\mathrm{eff}(2\omega)|^2
\end{equation}
where $\bm{e}(2\omega)$ is the unit vector denoting the direction of the analyzer collecting the reflected SHG beam. In our crossed polarization channel, the SHG analyzer is maintained perpendicular to the incident fundamental beam $\bm{e}(2\omega)\perp \bm{E}(\omega)$.

In the frame of the crystal, the non-zero independent elements of the $\chi^\mathrm{EQ}_{ijkl}$ tensor are determined by applying symmetries in the $S_6$ point group and SH permutation symmetry between indices $j$ and $l$ to the tensors. This yields only 18 non-vanishing independent tensor elements \cite{RN3}
\begin{gather*}
    \chi_{xxxx} = \chi_{yyyy} = \chi_{xxyy} + \chi_{xyyx} + \chi_{xyxy} \\
    \chi_{xxyy} = \chi_{yyxx} = \chi_{xyyx} = \chi_{yxxy} \\
    \chi_{xyxy} = \chi_{yxyx} \\
    \chi_{yyzz} = \chi_{xxzz} = \chi_{yzzy} = \chi_{xzzx} \\
    \chi_{zzyy} = \chi_{zzxx} = \chi_{zyyz} = \chi_{zxxz} \\
    \chi_{yzyz} = \chi_{xzxz} \\
    \chi_{zyzy} = \chi_{zxzx} \\
    \chi_{xyzz} = -\chi_{yxzz} = \chi_{xzzy} = -\chi_{yzzx} \\
    \chi_{zzxy} = -\chi_{zzyx} = \chi_{zyxz} = -\chi_{zxyz} \\
    \chi_{xzyz} = -\chi_{yzxz}
\end{gather*}

\begin{gather*}
    \begin{split}
        \chi_{xxxy} & = \chi_{xyxx} = -\chi_{yyyx} = -\chi_{yxyy} \\
                    & = \chi_{yyxy} + \chi_{yxyy} + \chi_{xyyy}
    \end{split}\\
    \chi_{yyxy} = -\chi_{xxyx} \\
    \chi_{xyyy} = -\chi_{yxxx} \\
    \begin{split}
        & \chi_{yyyz} = -\chi_{yxxz} = -\chi_{xyxz} = -\chi_{xxyz} \\
        = & \chi_{yzyy} = -\chi_{xzyx} = -\chi_{xzxy} = -\chi_{yzxx} 
    \end{split}\\
    \begin{split}
        & \chi_{xxxz} = -\chi_{xyyz} = -\chi_{yxyz} = -\chi_{yyxz} \\
        = & \chi_{xzxx} = -\chi_{xzyy} = -\chi_{yzxy} = -\chi_{yzyx} 
    \end{split}\\
    \chi_{xxzx} = -\chi_{xyzy} = -\chi_{yxzy} = -\chi_{yyzx} \\
    \chi_{zxxx} = -\chi_{zxyy} = -\chi_{zyxy} = -\chi_{zyyx} \\
    \chi_{yyzy} = -\chi_{yxzx} = -\chi_{xyzx} = -\chi_{xxzy} \\
    \chi_{zyyy} = -\chi_{zyxx} = -\chi_{zxyx} = -\chi_{zxxy} \\
    \chi_{zzzz}
\end{gather*}

With normal incidence on the (001) surface of the crystal, only two tensor elements survive in Eq. (\ref{SHGInt})
\begin{equation}\label{SP}
    \begin{split}
        I^{2\omega}_\perp(\varphi) & =\big|\chi_{xxzx}\sin3\varphi + \chi_{yyzy}\cos3\varphi\big|^2 \\
        & = \big|\chi_{xxzx}^2 + \chi_{yyzy}^2\big|\cos3\Big[\varphi-\frac{1}{3}\mathrm{atan}\frac{\chi_{xxzx}}{\chi_{yyzy}}\Big]
    \end{split}
\end{equation}
All the RA-SHG patterns in the main text have been fitted to this expression to obtain $A=|\chi_{xxzx}^2+\chi_{yyzy}^2|$ and $\varphi_0=\frac{1}{3}\mathrm{atan}\frac{\chi_{xxzx}}{\chi_{yyzy}}$. From Eq. (\ref{SP}) it is noted that any changes in $\chi_{xxzx}$ and/or $\chi_{yyzy}$ will lead to the modification of RA-SHG amplitude $A$ and orientation $\varphi_0$.

\subsection{Time-resolved linear reflectivity measurements on the CCDW in \bm{$1T$}\text{-}\bm{$\mathrm{TaS}_2$}}

A reference time-resolved linear reflectivity measurement was conducted with 720 nm pump and 800 nm probe pulses. FIG. S3(a) and S3(b) show the trace and its corresponding FFT spectrum. The pump fluence for this measurement is set the same at 0.46 mJ/cm$^2$ as that in FIG. 3 in the main text, in which the FFT spectra from fitting RA-SHG amplitude and orientation traces are also reproduced here in FIG. S3(c) and S3(d) for comparison. In FIG. S3(b) we note an overall broad asymmetric mode peaked at around 2.38 THz where the triplet structure is not seen.

\begin{figure*}[h!]
\includegraphics[width=0.8\textwidth]{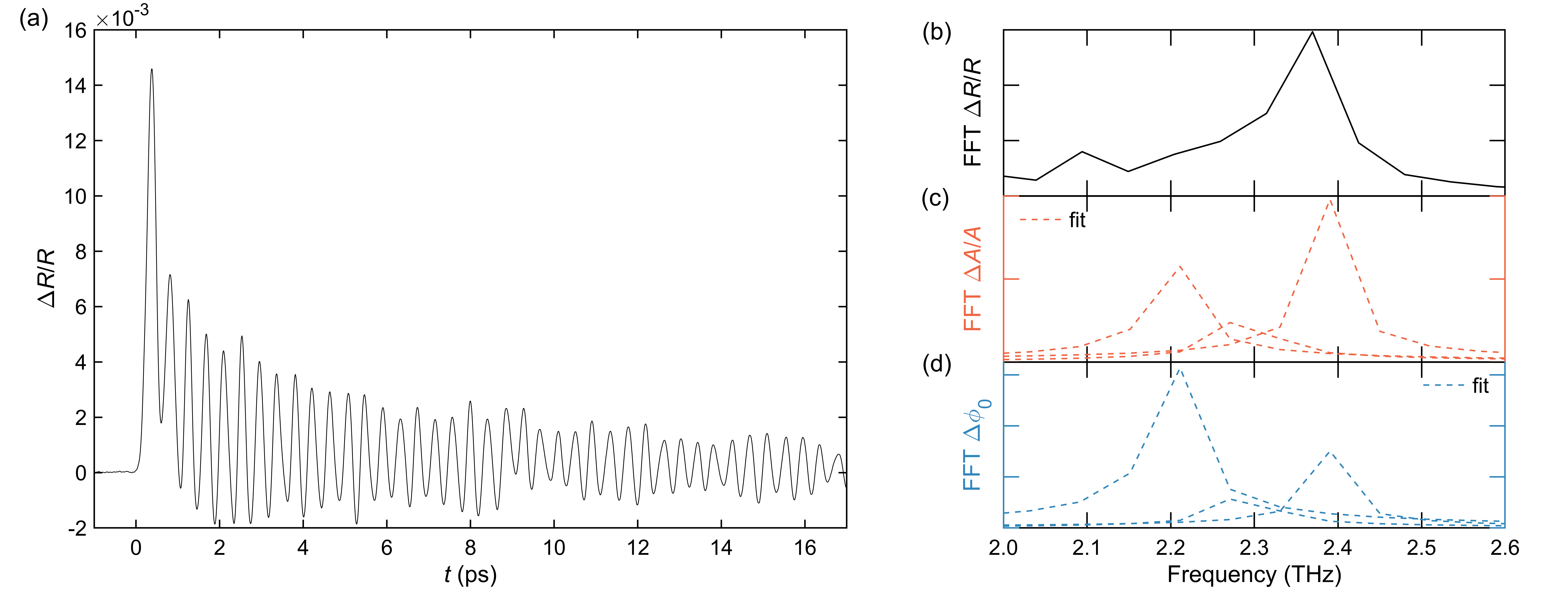}%
\caption{(a) The time-resolved linear reflectivity trace taken at 90 K in the CCDW phase of \sample{}. (b) The FFT spectrum of the trace in (a). (c-d) The FFT spectra of the RA-SHG amplitude and orientation traces. Data in (c) and (d) are reproduced from FIG. 3.}
\end{figure*}